\documentclass[amsmath,aps,showpacs,a4paper,10pt]{revtex4}

 \usepackage{epsf}
 \usepackage{graphicx}    

\begin{document}

 \newcommand{\be}[1]{\begin{equation}\label{#1}}
 \newcommand{\ee}{\end{equation}}
 \newcommand{\bea}{\begin{eqnarray}}
 \newcommand{\eea}{\end{eqnarray}}
 \def\disp{\displaystyle}

 \def\gsim{ \lower .75ex \hbox{$\sim$} \llap{\raise .27ex \hbox{$>$}} }
 \def\lsim{ \lower .75ex \hbox{$\sim$} \llap{\raise .27ex \hbox{$<$}} }

 \begin{titlepage}

 \begin{flushright}
 arXiv:0708.0884
 \end{flushright}

 \title{\Large \bf A New Model of Agegraphic Dark Energy}

 \author{Hao~Wei}
 \email[\,email address:\ ]{haowei@mail.tsinghua.edu.cn}
 \affiliation{Department of Physics and Tsinghua Center for
 Astrophysics,\\ Tsinghua University, Beijing 100084, China}

 \author{Rong-Gen Cai}
 \email[\,email address:\ ]{cairg@itp.ac.cn}
 \affiliation{Institute of Theoretical Physics, Chinese Academy
 of Sciences, P.O. Box 2735, Beijing 100080, China}

 \begin{abstract}\vspace{1cm}
 \centerline{\bf ABSTRACT}\vspace{2mm}
In this note, we propose a new model of agegraphic dark energy
 based on the K\'{a}rolyh\'{a}zy relation, where the time scale
 is chosen to be the conformal time $\eta$ of the
 Friedmann-Robertson-Walker~(FRW) universe. We find that in the
 radiation-dominated epoch, the equation-of-state parameter of
 the new agegraphic dark energy $w_q=-1/3$ whereas
 $\Omega_q=n^2a^2$; in the matter-dominated epoch, $w_q=-2/3$
 whereas $\Omega_q=n^2a^2/4$; eventually, the new agegraphic dark
 energy dominates; in the late time $w_q\to -1$ when $a\to\infty$,
 and the new agegraphic dark energy mimics a cosmological constant.
 In every stage, all things are consistent. The confusion in the
 original agegraphic dark energy model proposed in arXiv:0707.4049
 disappears in this new model. Furthermore, $\Omega_q\ll 1$ is
 naturally satisfied in both radiation-dominated and
 matter-dominated epochs where $a\ll 1$. In addition, we further
 extend the new agegraphic dark energy model by including the
 interaction between the new agegraphic dark energy and background
 matter. In this case, we find that $w_q$ can cross the phantom
 divide.
 \end{abstract}


 \maketitle

 \end{titlepage}

 \renewcommand{\baselinestretch}{1.6}



\section{Introduction}\label{sec1}
The cosmological constant problem is essentially a problem in
 quantum gravity, since the cosmological constant is commonly
 considered as the vacuum expectation value of some quantum
 fields. Before a completely successful quantum theory of
 gravity is available, it is more realistic to consider some
 consequences of combining quantum mechanics with general
 relativity directly.

In general relativity, one can measure the spacetime without
 any limit of accuracy. In quantum mechanics, however, the
 well-known Heisenberg uncertainty relation puts a limit of
 accuracy in these measurements. Following the line of quantum
 fluctuations of spacetime, K\'{a}rolyh\'{a}zy and his
 collaborators~\cite{r1} (see also~\cite{r2}) made an interesting
 observation concerning the distance measurement for Minkowski
 spacetime through a light-clock {\it Gedanken experiment},
 namely, the distance $t$ in Minkowski spacetime cannot be known
 to a better accuracy than
 \be{eq1}
 \delta t=\lambda t_p^{2/3}t^{1/3},
 \ee
 where $\lambda$ is a dimensionless constant of order unity. We use
 the units $\hbar=c=k_B=1$ throughout this work. Thus, one can use
 the terms like length and time interchangeably, whereas
 $l_p=t_p=1/m_p$ with $l_p$, $t_p$ and $m_p$ being the reduced
 Planck length, time and mass respectively.

The K\'{a}rolyh\'{a}zy relation~(\ref{eq1}) together with the
 time-energy uncertainty relation enables one to estimate a quantum
 energy density of the metric fluctuations of Minkowski
 spacetime~\cite{r3,r2}. Following~\cite{r3,r2}, with respect to
 Eq.~(\ref{eq1}) a length scale $t$ can be known with a maximum
 precision $\delta t$ determining thereby a minimal detectable cell
 $\delta t^3\sim t_p^2 t$ over a spatial region $t^3$. Such a cell
 represents a minimal detectable unit of spacetime over a given
 length scale $t$. If the age of the Minkowski spacetime is $t$,
 then over a spatial region with linear size $t$ (determining the
 maximal observable patch) there exists a minimal cell $\delta t^3$
 the energy of which due to time-energy uncertainty relation cannot
 be smaller than~\cite{r3,r2}
 \be{eq2}
 E_{\delta t^3}\sim t^{-1}.
 \ee
 Therefore, the energy density of metric fluctuations of
 Minkowski spacetime is given by~\cite{r3,r2}
 \be{eq3}
 \rho_q\sim\frac{E_{\delta t^3}}{\delta t^3}\sim
 \frac{1}{t_p^2 t^2}\sim\frac{m_p^2}{t^2}.
 \ee
 We refer to the original papers~\cite{r3,r2} for more details.
 It is worth noting that in fact, the K\'{a}rolyh\'{a}zy
 relation~(\ref{eq1}) and the corresponding energy
 density~(\ref{eq3}) have been independently rediscovered later
 for many times in the literature (see e.g.~\cite{r4,r5,r6}).

In~\cite{r3} (see also~\cite{r7}), it is noticed that the
 K\'{a}rolyh\'{a}zy relation~(\ref{eq1}) naturally obeys the
 holographic black hole entropy bound. In fact, the holographic
 dark energy~\cite{r8} also stems from  the idea of holographic
 black hole entropy bound~\cite{r9}. For a complete list of
 references concerning the holographic dark energy, one can see
 e.g.~\cite{r10} and references therein. It is worth noting that
 the form of energy density Eq.~(\ref{eq3}) is similar to the one
 of holographic dark energy~\cite{r8,r9,r10,r11,r12}, i.e.,
 $\rho_\Lambda\sim l_p^{-2}l^{-2}$. The similarity between
 $\rho_q$ and $\rho_\Lambda$ might reveal some universal features
 of quantum gravity, although they arise from different ways.
 See~\cite{r7} for a detailed discussion on this point.

In the next section, we briefly review the original agegraphic dark
 energy model proposed in~\cite{r7}. The difficulties of this
 original agegraphic dark energy model are also discussed. In
 Sec.~\ref{sec3}, we propose a new model of agegraphic dark
 energy and find that the confusion in the original agegraphic dark
 energy model does not exist in this new model. In Sec.~\ref{sec4},
 we further extend the new agegraphic dark energy model by including
 the interaction between the new agegraphic dark energy and the
 background matter. Some concluding remarks are given in
 Sec.~\ref{sec5}.


\section{The original agegraphic dark energy model}\label{sec2}
Based on the energy density~(\ref{eq3}), a so-called agegraphic dark
 energy model was proposed in~\cite{r7}. There, as the most natural
 choice, the time scale $t$ in Eq.~(\ref{eq3}) is chosen to
 be the age of the universe
 \be{eq4}
 T=\int_0^a\frac{da}{Ha},
 \ee
 where $a$ is the scale factor of our universe; $H\equiv\dot{a}/a$
 is the Hubble parameter; a dot denotes the derivative with respect
 to cosmic time. Thus, the energy density of the agegraphic dark
 energy is given by~\cite{r7}
 \be{eq5}
 \rho_q=\frac{3n^2m_p^2}{T^2},
 \ee
 where the numerical factor $3n^2$ is introduced to parameterize
 some uncertainties, such as the species of quantum fields in
 the universe, the effect of curved spacetime (since the energy
 density is derived for Minkowski spacetime), and so on. Obviously,
 since the present age of the universe $T_0\sim H_0^{-1}$ (the
 subscript ``0'' indicates the present value of the corresponding
 quantity; we set $a_0=1$), the present energy density of the
 agegraphic dark energy explicitly meets the observed value
 naturally, provided that the numerical factor $n$ is of order
 unity. In addition, by choosing the age of the universe rather than
 the future event horizon as the length measure, the drawback
 concerning causality in the holographic dark energy model~\cite{r8}
 does not exist in the agegraphic dark energy model~\cite{r7}.

If we consider a flat Friedmann-Robertson-Walker~(FRW) universe
 containing the agegraphic dark energy and pressureless matter, the
 corresponding Friedmann equation reads
 \be{eq6}
 H^2=\frac{1}{3m_p^2}\left(\rho_m+\rho_q\right).
 \ee
 It is convenient to introduce the fractional energy densities
 $\Omega_i\equiv\rho_i/(3m_p^2H^2)$ for $i=m$ and $q$. From
 Eq.~(\ref{eq5}), it is easy to find that
 \be{eq7}
 \Omega_q=\frac{n^2}{H^2T^2},
 \ee
 whereas $\Omega_m=1-\Omega_q$ from Eq.~(\ref{eq6}). By using
 Eqs.~(\ref{eq5})---(\ref{eq7}) and the energy conservation
 equation $\dot{\rho}_m+3H\rho_m=0$, we obtain the equation
 of motion for $\Omega_q$ as~\cite{r7}
 \be{eq8}
 \Omega_q^\prime=\Omega_q\left(1-\Omega_q\right)
 \left(3-\frac{2}{n}\sqrt{\Omega_q}\right),
 \ee
 where a prime denotes the derivative with respect to the
 $e$-folding time $N\equiv\ln a$. From the energy conservation
 equation $\dot{\rho}_q+3H(\rho_q+p_q)=0$, as well as
 Eqs.~(\ref{eq5}) and~(\ref{eq7}), it is easy to find that the
 equation-of-state parameter~(EoS) of the agegraphic dark energy
 $w_q\equiv p_q/\rho_q$ is given by~\cite{r7}
 \be{eq9}
 w_q=-1+\frac{2}{3n}\sqrt{\Omega_q}.
 \ee

The agegraphic dark energy has some interesting features.
 From Eq.~(\ref{eq9}), it is easy to see that $w_q\to -1$ in the
 early time where $\Omega_q\to 0$, whereas $w_q\to -1+2/(3n)$ in
 the late time where $\Omega_q\to 1$. On the other hand, if one
 considers the matter-dominated epoch where $\Omega_q\ll 1$ while
 $a\ll 1$, Eq.~(\ref{eq8}) approximately becomes
 $\Omega_q^\prime\simeq 3\Omega_q$. Therefore, $\Omega_q\propto a^3$
 in the matter-dominated epoch. This is consistent with the fact
 that in the matter-dominated epoch the agegraphic dark energy
 mimics a cosmological constant, $w_q\simeq -1$, whereas the energy
 density of pressureless matter scales as $\rho_m\propto a^{-3}$.

However, there is an implicit confusion in this agegraphic dark
 energy model~\cite{r13}. Actually, in the matter-dominated epoch
 with $\Omega_q\ll 1$, one has $a\propto t^{2/3}$, thus
 $T^2\propto a^3$. From Eq.~(\ref{eq5}), in this epoch
 $\rho_q\propto a^{-3}$. Since $\rho_m\propto a^{-3}$, one has
 $\Omega_q\simeq const.$, which is conflict with
 $\Omega_q\propto a^3$ obtained previously. In addition, from
 Eq.~(\ref{eq5}), the agegraphic dark energy tracks the dominated
 components (either pressureless matter or radiation). Therefore,
 the agegraphic dark energy never dominates. This is of course
 unacceptable.

To get around this confusion, there are two different solutions. The
 first one is to replace $T$ with $T+\delta$, where $\delta$ is a
 constant with dimension of time. Thus, Eq.~(\ref{eq5})
 becomes $\rho_q=3n^2m_p^2(T+\delta)^{-2}$. In the early time,
 $T\ll\delta$, thus $\rho_q\simeq const.$, i.e., the agegraphic dark
 energy mimics a cosmological constant. In the late time,
 $T\gg\delta$, thus $\rho_q$ can be described approximately by
 Eq.~(\ref{eq5}). In the intermediate stage $T\sim\delta$,
 $\delta$ cannot be neglected. So, the tracking behavior no
 longer exists. However, this solution abandons the motivation of
 the K\'{a}rolyh\'{a}zy relation~(\ref{eq1}), and then becomes only
 a phenomenological model.

The second solution argues that Eq.~(\ref{eq5}) was derived for the
 Minkowski spacetime and is not valid exactly in the early time
 where the spacetime is highly curved. Thus, $n=n(T)$ may be variable
 in the early time, say, $n\propto T$. Only after a critical $T_c$,
 the factor $n$ can be approximated to a constant in Eq.~(\ref{eq5}).
 One can choose $T_c$ in the early or middle stage of the
 matter-dominated epoch. For $T>T_c$, the energy density of
 agegraphic dark energy cannot be ignored already (nb.
 $\Omega_q\sim 0.1-0.2$ in the matter-dominated epoch does not
 violate the constraint from the BBN~\cite{r14}). Therefore, the
 tracking behavior no longer exists, and then the confusion
 disappears. However, this solution is successful at the price of
 giving up the validity of Eq.~(\ref{eq5}) in the early time.


\section{A new model of agegraphic dark energy}\label{sec3}
In this note, we seek another more comfortable solution to the
 confusion mentioned above. The history always repeats itself
 (``there is nothing new under the sun'', Ecclesiastes 1:9).
 It is suggestive to review the history of the holographic
 dark energy model~\cite{r8,r9,r10,r11,r12}. In~\cite{r9},
 Cohen {\it et al.} argued that $\rho_\Lambda=3c^2m_p^2L^{-2}$
 from the holographic principle. The most natural choice of $L$
 is the Hubble horizon $H^{-1}$. However, Hsu~\cite{r11} (see
 also~\cite{r8}) pointed out that in this case the EoS of
 holographic dark energy is zero and the expansion of the universe
 cannot be accelerated. The next choice of $L$ is the particle
 horizon. Unfortunately, in this case, the EoS of holographic dark
 energy is always larger than $-1/3$ and the expansion of the
 universe also cannot be accelerated~\cite{r8}. Finally,
 Li~\cite{r8} found out that $L$ might be the future
 event horizon of the universe.
 In this case, the EoS of holographic dark energy
 is given by~\cite{r8,r12}
 \be{eq10}
 w_\Lambda=-\frac{1}{3}-\frac{2}{3c}\sqrt{\Omega_\Lambda}.
 \ee
 Obviously, $w_\Lambda<-1/3$ and the expansion of the universe can
 indeed be accelerated. In fact, this is just the holographic dark
 energy model investigated extensively in the literature, although
 it has the drawback concerning causality, because the existence of
 the future event horizon requires an eternal accelerated expansion
 of the universe.

The development of the holographic dark energy model might shed
 some light on the agegraphic dark energy model. Can we have another
 choice of the time scale in Eq.~(\ref{eq3}) rather than $T$, the
 age of the universe?  The answer is yes. Here we choose the time
 scale to be the conformal time $\eta$ instead, which is defined by
 $dt=a\,d\eta$ [where $t$ is the cosmic time, do not confuse it with
 the $t$ in Eq.~(\ref{eq3})]. Notice that the K\'{a}rolyh\'{a}zy
 relation~(\ref{eq1}) is derived for Minkowski spacetime
 $ds^2=dt^2-d{\bf x}^2$~\cite{r1,r2,r3}. For the FRW universe,
 $ds^2=dt^2-a^2d{\bf x}^2=a^2(d\eta^2-d{\bf x}^2)$.
 Thus, it might be more reasonable to choose the time scale
 in Eq.~(\ref{eq3}) to be the conformal time $\eta$ since it is the
 causal time in the Penrose diagram of the FRW universe~\cite{r17}.
 Then, we propose the new agegraphic dark energy as
 \be{eq11}
 \rho_q=\frac{3n^2m_p^2}{\eta^2},
 \ee
 where the conformal time
 \be{eq12}
 \eta\equiv\int\frac{dt}{a}=\int\frac{da}{a^2H}.
 \ee
 If we write $\eta$ to be a definite integral, there will be an
 additional integration constant. Thus, $\dot{\eta}=1/a$. The
 corresponding fractional energy density is given by
 \be{eq13}
 \Omega_q=\frac{n^2}{H^2\eta^2}.
 \ee

At first, we consider a flat FRW universe containing the new
 agegraphic dark energy and pressureless matter. By using
 Eqs.~(\ref{eq6}), (\ref{eq13}), (\ref{eq11}) and the energy
 conservation equation $\dot{\rho}_m+3H\rho_m=0$, we find that
 the equation of motion for $\Omega_q$ is given by
 \be{eq14}
 \frac{d\Omega_q}{da}=\frac{\Omega_q}{a}\left(1-\Omega_q\right)
 \left(3-\frac{2}{n}\frac{\sqrt{\Omega_q}}{a}\right).
 \ee
 From the energy conservation equation
 $\dot{\rho}_q+3H(\rho_q+p_q)=0$, as well as Eqs.~(\ref{eq13})
 and~(\ref{eq11}), it is easy to find that the EoS of new
 agegraphic dark energy $w_q\equiv p_q/\rho_q$ is given by
 \be{eq15}
 w_q=-1+\frac{2}{3n}\frac{\sqrt{\Omega_q}}{a}.
 \ee
 Comparing with Eqs.~(\ref{eq8}) and~(\ref{eq9}), the scale factor
 $a$ enters Eqs.~(\ref{eq14}) and~(\ref{eq15}) explicitly. When
 $a\to\infty$, $\Omega_q\to 1$, thus $w_q\to -1$ in the late time.
 When $a\to 0$, $\Omega_q\to 0$, we cannot obtain $w_q$ from
 Eq.~(\ref{eq15}) directly. Let us consider the matter-dominated
 epoch, $H^2\propto\rho_m\propto a^{-3}$. Thus,
 $a^{1/2}da\propto dt=ad\eta$. Therefore, $\eta\propto a^{1/2}$.
 From Eq.~(\ref{eq11}), $\rho_q\propto a^{-1}$.  From the energy
 conservation equation $\dot{\rho}_q+3H\rho_q(1+w_q)=0$, we obtain
 that $w_q=-2/3$. Since $\rho_m\propto a^{-3}$, it is expected that
 $\Omega_q\propto a^2$. Comparing $w_q=-2/3$ with Eq.~(\ref{eq15}),
 we find that $\Omega_q=n^2a^2/4$ in the matter-dominated epoch
 as expected. For $a\ll 1$, if $n$ is of order unity,
 $\Omega_q\ll 1$ naturally. On the other hand, one can check that
 $\Omega_q=n^2a^2/4$ satisfies
 $$\frac{d\Omega_q}{da}=\frac{\Omega_q}{a}
 \left(3-\frac{2}{n}\frac{\sqrt{\Omega_q}}{a}\right),$$
 which is the approximation of Eq.~(\ref{eq14}) for
 $1-\Omega_q\simeq 1$. So, all things are consistent. The confusion
 in the original agegraphic dark energy model~\cite{r7} does not
 exist in this new model.

In fact, we can extend our discussion to include the
 radiation-dominated epoch. To be general, we consider a flat FRW
 universe containing the new agegraphic dark energy and background
 matter whose EoS is constant $w_m$. In particular, $w_m=0$ for
 pressureless matter, whereas $w_m=1/3$ for radiation. By using
 Eqs.~(\ref{eq6}), (\ref{eq13}), (\ref{eq11}) and the energy
 conservation equation $\dot{\rho}_m+3H\rho_m(1+w_m)=0$, we find
 that the equation of motion for $\Omega_q$ is given by
 \be{eq16}
 \frac{d\Omega_q}{da}=\frac{\Omega_q}{a}\left(1-\Omega_q\right)
 \left[3\left(1+w_m\right)-\frac{2}{n}\frac{\sqrt{\Omega_q}}{a}\right].
 \ee
 On the other hand, the EoS of new agegraphic dark energy is the same
 one given in Eq.~(\ref{eq15}). In the radiation-dominated epoch,
 $H^2\propto\rho_r\propto a^{-4}$. Thus, $ada\propto dt=ad\eta$.
 Therefore, $\eta\propto a$. From Eq.~(\ref{eq11}),
 $\rho_q\propto a^{-2}$. From the energy conservation
 equation $\dot{\rho}_q+3H\rho_q(1+w_q)=0$, we obtain that $w_q=-1/3$
 in the radiation-dominated epoch. Since $\rho_r\propto a^{-4}$, it is
 expected that $\Omega_q\propto a^2$. Comparing $w_q=-1/3$ with
 Eq.~(\ref{eq15}), we find that $\Omega_q=n^2a^2$ in the
 radiation-dominated epoch as expected. For $a\ll 1$, if $n$ is of
 order unity, $\Omega_q\ll 1$ naturally. On the other hand, one can
 check that $\Omega_q=n^2a^2$ satisfies
 $$\frac{d\Omega_q}{da}=\frac{\Omega_q}{a}
 \left(4-\frac{2}{n}\frac{\sqrt{\Omega_q}}{a}\right),$$
 which is the approximation of Eq.~(\ref{eq16}) for
 $1-\Omega_q\simeq 1$ and $w_m=1/3$. Once again, we see that all
 things are consistent in the radiation-dominated epoch.

In summary, in the radiation-dominated epoch, $w_q=-1/3$ whereas
 $\Omega_q=n^2a^2$; in the matter-dominated epoch, $w_q=-2/3$
 whereas $\Omega_q=n^2a^2/4$; eventually, the new agegraphic dark
 energy dominates; in the late time $w_q\to -1$ when $a\to\infty$,
 the new agegraphic dark energy mimics a cosmological constant.
 It is worth noting that $\Omega_q\ll 1$ naturally in both
 radiation-dominated and matter-dominated epochs where $a\ll 1$.

It is easy to see that the behavior of new agegraphic dark
 energy is very different from the one of original agegraphic
 dark energy proposed in~\cite{r7}. In addition, we notice that
 the evolution behavior of new agegraphic dark energy is similar
 to that of holographic dark energy~\cite{r8,r9,r10,r11,r12}.
 From Eq.~(\ref{eq10}), the EoS of holographic dark energy
 $w_\Lambda\simeq -1/3$ in the radiation-dominated and
 matter-dominated epochs where $\Omega_\Lambda\ll 1$, whereas
 $w_\Lambda\to -1/3-2/(3c)$ in the late time $a\to\infty$ where
 $\Omega_\Lambda\to 1$. If $c=1$, the holographic dark energy
 mimics a cosmological constant in the late time. However, for
 the new agegraphic dark energy, $w_q=-2/3$ in the matter-dominated
 epoch; $w_q\to -1$ in the late time regardless of the value of
 $n$. As a result, there  also exist some essential differences
 between the new agegraphic dark energy and the holographic dark
 energy.


\section{The new agegraphic dark energy model with interaction}\label{sec4}
Following~\cite{r15,r16}, we further extend the new agegraphic
 dark energy model by including the interaction between the new
 agegraphic dark energy and background matter whose EoS is
 constant $w_m$. We assume that the new agegraphic dark energy and
 background matter exchange energy through an interaction term $Q$
 as
 \bea
 &&\dot{\rho}_q+3H\rho_q\left(1+w_q\right)=-Q,\label{eq17}\\
 &&\dot{\rho}_m+3H\rho_m\left(1+w_m\right)=Q.\label{eq18}
 \eea
 In this way the total energy conservation equation
 $\dot{\rho}_{tot}+3H\left(\rho_{tot}+p_{tot}\right)=0$ is still kept. In this
 case, by using Eqs.~(\ref{eq6}), (\ref{eq13}),
 (\ref{eq11}) and~(\ref{eq18}), we find that the equation of
 motion for $\Omega_q$ is changed to
 \be{eq19}
 \frac{d\Omega_q}{da}=\frac{\Omega_q}{a}
 \left\{\left(1-\Omega_q\right)\left[3\left(1+w_m\right)
 -\frac{2}{n}\frac{\sqrt{\Omega_q}}{a}\right]
 -\frac{Q}{3m_p^2H^3}\right\}.
 \ee
 From Eqs.~(\ref{eq17}), (\ref{eq13}) and~(\ref{eq11}), we obtain
 the EoS of new agegraphic dark energy
 \be{eq20}
 w_q=-1+\frac{2}{3n}\frac{\sqrt{\Omega_q}}{a}-\frac{Q}{3H\rho_q}.
 \ee
 It is easy to see that Eqs.~(\ref{eq19}) and~(\ref{eq20}) reduce
 to Eqs.~(\ref{eq16}) and~(\ref{eq15}) in the case of $Q=0$ (i.e.
 without interaction). It is worth noting that from Eq.~(\ref{eq15})
 $w_q$ is always larger than $-1$ and cannot cross the phantom
 divide $w=-1$ in the case of $Q=0$ (i.e. without interaction).
 Here, the situation is changed by the interaction $Q\not=0$.
 If $Q>0$, from Eq.~(\ref{eq20}), one can see that $w_q$
 can be smaller than $-1$ or larger than $-1$. Thus, it becomes
 possible that $w_q$ crosses the phantom divide.


\section{Concluding remarks}\label{sec5}
In this note, we propose a new model of agegraphic dark energy
 based on the K\'{a}rolyh\'{a}zy relation~(\ref{eq1}), where
 the time scale is chosen to be the conformal time $\eta$ of
 the FRW universe.  We find in this model that in the
 radiation-dominated epoch, the EoS of new agegraphic dark energy
 $w_q=-1/3$ whereas $\Omega_q=n^2a^2$; in the matter-dominated
 epoch, $w_q=-2/3$ whereas $\Omega_q=n^2a^2/4$; eventually, the
 new agegraphic dark energy dominates; in the late time
 $w_q\to -1$ when $a\to\infty$, and the new agegraphic dark energy
 mimics a cosmological constant. In every stage, all things are
 consistent. The confusion in the original agegraphic dark energy
 model proposed in~\cite{r7} disappears in this model. Furthermore,
 $\Omega_q\ll 1$ naturally occurs in both radiation-dominated and
 matter-dominated epochs where $a\ll 1$. In addition, we further
 study the new agegraphic dark energy model by including the
 interaction between the new agegraphic dark energy and background
 matter. In this case, we find that $w_q$ can cross the phantom
 divide.

The evolution behavior of the new agegraphic dark energy is very
 different from that of original agegraphic dark energy proposed
 in~\cite{r7}. Instead the evolution behavior of the new
 agegraphic dark energy is similar to that of the holographic dark
 energy~\cite{r8,r9,r10,r11,r12}. But, as mentioned above, some
 essential differences exist between them.  In particular, the new
 agegraphic dark energy model is free of the drawback concerning
 causality problem which exists in the holographic dark energy
 model~\cite{r8}. Another advantage of the new agegraphic dark
 energy model is that the K\'{a}rolyh\'{a}zy relation~(\ref{eq1})
 naturally obeys the holographic black hole entropy
 bound~\cite{r3,r7}.

Furthermore, thanks to its special
 analytic features in the radiation-dominated and matter-dominated
 epochs, this new agegraphic dark energy model is actually a
 {\em single-parameter} model~\cite{r18}. To our knowledge, it is
 the third {\em single-parameter} cosmological model besides the
 well-known $\Lambda$CDM model and the DGP braneworld
 model~\cite{r19}. Also, it is found that the coincidence problem
 can be solved naturally in this model~\cite{r18}. In addition, as
 shown in~\cite{r18}, the new agegraphic dark energy model can fit
 the cosmological observations of type Ia supernovae, cosmic
 microwave background, and large scale structure very well. In this
 sense, it is expected that the new agegraphic dark energy model
 is a cosmologically viable model for dark energy.

Finally, it is worth noting that there are still some problems in
 the (new) agegraphic dark energy model. After the appearance of
 our relevant works on the (new) agegraphic dark energy, it is
 found that the original agegraphic dark energy model proposed
 in~\cite{r7,r15,r16} is difficult to reconcile with the
 big bang nucleosynthesis~(BBN) constraint~\cite{r20}. On the
 other hand, as shown in~\cite{r21}, the situation is better
 in the new agegraphic dark energy model. To alleviate the
 difficulty from BBN constraint, one can introduce the coupling
 between the dark matter and dark energy (for references on this
 topic, see e.g.~\cite{r15,r16,r22} and references therein).
 In addition, the (new) agegraphic dark energy model faces the
 problem of instabilities~\cite{r23}, while the holographic dark
 energy model also faces the same problem~\cite{r24}. Nevertheless,
 we consider that the new agegraphic dark energy model deserves
 further investigations, since it has some valuable advantages
 mentioned above. We hope that it can shed new light on the
 understanding of the mysterious dark energy.


\section*{ACKNOWLEDGMENTS}
We are grateful to Prof.~Miao~Li and Prof.~Shuang~Nan~Zhang for
 helpful discussions. We also thank Minzi~Feng, as well as Hui~Li,
 Yi~Zhang, Xing~Wu, Li-Ming~Cao, Xin~Zhang, Jian~Wang, Bin~Hu for
 useful discussions. This work was supported in part by a grant
 from China Postdoctoral Science Foundation, a grant from Chinese
 Academy of Sciences~(No.~KJCX3-SYW-N2), and by NSFC under grants
 No.~10325525, No.~10525060 and No.~90403029.

\renewcommand{\baselinestretch}{1.2}


\end{document}